\documentclass[sigconf,nonacm,screen]{acmart}

\usepackage{prelude}

\usepackage{iftex}
\ifxetex
  \newfontfamily\redemonfont{TsukimiRounded}[
    Path = Tsukimi_Rounded/,
    Extension = .ttf,
    Ligatures = TeX,
    Scale = MatchUppercase,
    UprightFont = *-Medium,
    BoldFont = *-Bold,
  ]
  
  \newcommand\Rdemon{{\redemonfont ReDemon UI}}
  \newcommand\Sketch{{\redemonfont Sketch}}
  \newcommand\Synthesized{{\redemonfont Synthesized}}
  \newcommand\Timelines{{\redemonfont Timelines}}
\else\ifluatex
  \newfontfamily\redemonfont{TsukimiRounded}[
    Path = Tsukimi_Rounded/,
    Extension = .ttf,
    Ligatures = TeX,
    Scale = MatchUppercase,
    UprightFont = *-Medium,
    BoldFont = *-Bold,
  ]
  
  \newcommand\Rdemon{{\redemonfont ReDemon UI}}
  \newcommand\Sketch{{\redemonfont Sketch}}
  \newcommand\Synthesized{{\redemonfont Synthesized}}
  \newcommand\Timelines{{\redemonfont Timelines}}
\else
  \newcommand\Rdemon{\textsf{ReDemon UI}}
  \newcommand\Sketch{\textsf{Sketch}}
  \newcommand\Synthesized{\textsf{Synthesized}}
  \newcommand\Timelines{\textsf{Timelines}}
\fi

\let\origfootnote\footnote
\renewcommand{\footnote}[1]{\kern.06em\origfootnote{#1}}
\newcommand{\pfootnote}[1]{\kern-.06em\origfootnote{#1}}

\begin{document}
\title{\Rdemon: Reactive Synthesis by Demonstration for Web UI}

\author{Jay Lee}
\orcid{0000-0002-2224-4861}
\email{jhlee@ropas.snu.ac.kr}
\affiliation{%
  \institution{Seoul National University}
  \city{Seoul}
  \country{Korea}}

\author{Gyuhyeok Oh}
\orcid{0009-0008-4693-5377}
\email{ghoh@ropas.snu.ac.kr}
\affiliation{%
  \institution{Seoul National University}
  \city{Seoul}
  \country{Korea}}
  
\author{Joongwon Ahn}
\orcid{0009-0007-0055-7297}
\email{jwahn@ropas.snu.ac.kr}
\affiliation{%
  \institution{Seoul National University}
  \city{Seoul}
  \country{Korea}}

\author{Xiaokang Qiu}
\orcid{0000-0001-9476-7349}
\email{xkqiu@purdue.edu}
\affiliation{%
  \institution{Purdue University}
  \city{West Lafayette}
  \state{IN}
  \country{USA}}

\keywords{React, Programming by demonstration, Program synthesis, Sketch-based synthesis, Large language models}

\ccsdesc[500]{Human-centered computing~User interface programming}
\ccsdesc[500]{Human-centered computing~Graphical user interfaces}
\ccsdesc[500]{Software and its engineering~Programming by example}
\ccsdesc[500]{Software and its engineering~Automatic programming}

\begin{abstract}
  \Rdemon{} synthesizes React applications from user demonstrations, enabling designers and non-expert programmers to create UIs that integrate with standard UI prototyping workflows.
  Users provide a static mockup sketch with event handler holes and demonstrate desired runtime behaviors by interacting with the rendered mockup and editing the sketch.
  \Rdemon{} identifies reactive data and synthesizes a React program with correct state update logic.
  We utilize enumerative synthesis for simple UIs and LLMs for more complex UIs.
\end{abstract}

\begin{teaserfigure}
  \centering
  \includegraphics[width=\textwidth]{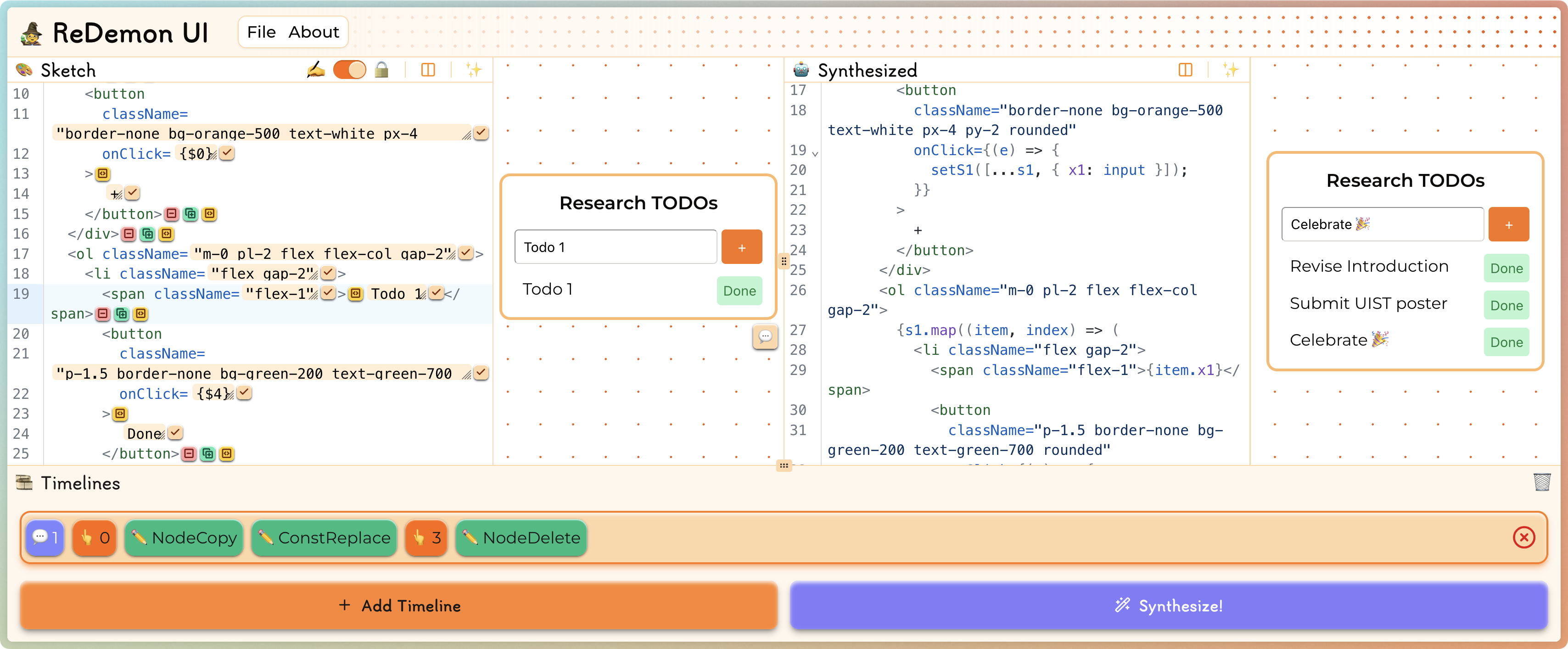}
  \caption{A screenshot of \Rdemon{} synthesizing a TODO list program (top-right pane) from user demonstrations (bottom pane) by interacting with the sketch (top-left pane). }\label{fig:teaser}
\end{teaserfigure}

\maketitle

\section{Introduction}
Creating interactive graphical applications is a complex process that involves dealing with dynamic behaviors and hierarchical user interfaces (UIs), requiring significant programming expertise.
Non-experts often struggle with this complexity, even when equipped with visual or prototyping tools that typically require users to manually implement reactivity.

Recent advancements in large language models (LLMs) have led to a proliferation of no-code and low-code platforms such as Stitch (\href{https://stitch.withgoogle.com/}{stitch.withgoogle.com}), Anima (\href{https://www.animaapp.com/}{animaapp.com}),%
and Uizard (\href{https://uizard.io/}{uizard.io}), allowing for rapid prototyping of UIs.
These tools enable even non-programmers to create interactive web applications.

However, tools that generate UIs with LLMs suffer from notable limitations~\citep{AhmImr25Role,WeiCouLamDraMaa25AIInspired,PatGagSheThaVarShi25StarryStudioAI}: (1)~heavy dependence on natural language prompts, (2)~generated code that may not be accurate or reliable, and (3)~difficulty in customizing application behavior.
Lacking fine-grained methods to describe intent, current approaches often lead to endless iterations of prompting to achieve the desired result~\citep{ZamWeiXiaGuJunLeeHarYan23Herding}.

We have developed \Rdemon{} (\cref{fig:teaser}), a system for synthesizing reactive user interfaces in React (\href{https://react.dev/}{react.dev}) directly from user demonstrations and a static mockup.
\Rdemon{} serves as a viable substitute for current prompt-based tools due to the following salient features:
\begin{enumerate}
  \item By adopting programming-by-demonstration~(PBD)~\citep{CypHalKurLieMauMyeTur93Watch}, \Rdemon{} enables users to directly express intent through interaction sequences rather than natural language prompts.
  \item \Rdemon{} automatically extracts reactive parameters from user demonstrations and synthesizes stateful React components with correct state management logic.
  \item Users can interactively provide additional demonstrations if the output does not match their intent, until a desirable UI is produced.
\end{enumerate}

\section{\Rdemon{} Frontend Interface}\label{sec:frontend}
Our frontend interface (\cref{fig:teaser}) consists of three panes:
(1)~the top-left \textbf{\Sketch{} pane}, where users provide a static UI mockup;
(2)~the top-right \textbf{\Synthesized{} pane}, which displays the synthesized React program; and
(3)~the bottom \textbf{\Timelines{} pane}, which shows recorded demonstrations.

In the \Sketch{} pane, users provide JSX~\citep{Met22JSX} annotated with \verb/$/-prefixed numbered holes (e.g., \verb/$1/, \verb/$2/) for event handlers.
\Rdemon{} recognizes these holes and follows a \emph{sketch}-based synthesis~\citep{SolTanBodSesSar06Combinatorial} to fill in the event handlers.
The sketch can be seen as a partial specification of the UI, representing the static structure for which the synthesizer implements the interactive logic.

When users are ready to demonstrate how the UI should behave at runtime, they can lock the sketch to enter \emph{demo mode.}

\begin{figure}[t]
  \centering
  \includegraphics[width=\linewidth]{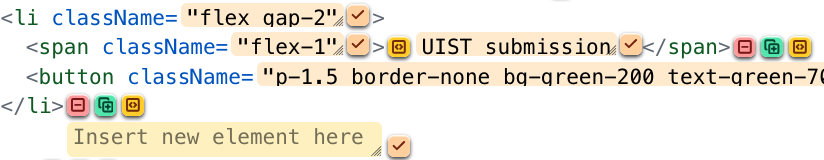}
  \caption{In demo mode, users can structurally edit the sketch to demonstrate the runtime behaviors of the UI.}\label{fig:edit}
\end{figure}

In demo mode, \Rdemon{} records \emph{actions} applied to the rendered mockup while presenting widgets in the editor to structurally \emph{edit} the sketch (\cref{fig:edit}).
\Rdemon{} currently supports button clicks and text input actions.
An action can be followed by \emph{edits:} string value replacements, node deletions, copies, or insertions.

The \Timelines{} pane contains \emph{demo timelines} that record all actions and edits.
Each demo timeline represents a single sequence of user interaction.
Users can add more timelines as necessary and synthesize the React component using the two floating buttons at the bottom of the \Timelines{} pane.

The \Synthesized{} pane displays the synthesized stateful React component generated from the sketch and demo timelines.
When the synthesized output does not match the user's intent---either due to underspecified demo timelines or LLM hallucinations---users can easily amend and append to demo timelines and re-synthesize.

\section{\Rdemon{} Synthesis Backend}\label{sec:backend}
\Rdemon{} first tries an enumerative method, which currently supports simple UIs (e.g., a counter) and falls back to LLM (Gemini 2.0 Flash) for more complex UIs (e.g., a TODO list shown in \cref{fig:teaser}).

Before calling the synthesis backend, \Rdemon{} algorithmically extracts reactive parameters from demo timelines.
By computing diffs of the UI from edits grouped by actions, \Rdemon{} identifies the reactive parts of the UI.
The sketch and demo timelines are then elaborated with these reactive parameters.

The enumerative method associates the reactive parameters with actions to synthesize candidate functions for event handler holes.

If the enumerative method fails, \Rdemon{} converts the elaborated sketch and demo timelines---capturing temporal changes in reactive state variables---into a textual format, which is provided to an LLM.
As a prompt, we provide a simple example and a desired output, which constitutes a form of inference-time one-shot learning~\citep{BroManRydSubKapDhaNeeShySasAskAgaHerKruHenChiRamZieWuWinHesCheSigLitGraCheClaBerMcCRadSutAmoLanguage}.
We have found that LLM is effective when latent states are needed to explain the demo timelines.


\section{Related Work}\label{sec:related}
\paragraph{Programming Interfaces by Demonstration}
The closest work to \Rdemon{} is a 35-year-old system, Peridot~\citep{Mye90Creating} and its successor Garnet~\citep{MyeGiuDanZanKosPerMicMar90Garnet}.
Peridot allows designers to draw UIs, automatically identifies constraints, and generates LISP code for rendering the UI.
However, Peridot primarily addressed low-level aspects of UI construction that are largely obsolete today.

There has been work on adapting UIs~\citep{NicLau08Mobilization} or identifying relevant UI code~\citep{OneMye09FireCrystal} by demonstration.
Highlight~\citep{NicLau08Mobilization} allows users to create mobile applications by clipping portions of existing desktop web pages and recording user interactions.
FireCrystal~\citep{OneMye09FireCrystal} helps developers understand the implementation of existing web pages by interacting with them.

\paragraph{UI Synthesis using Language Models}
Prior work~\citep{WeiCouLamDraMaa25AIInspired,XiaoChenLiChenSunZhou24Prototype2Code,PatGagSheThaVarShi25StarryStudioAI} mostly focused on generating static UIs using LLMs to interpret prompts, UI designs from tools like Figma (\href{https://www.figma.com/}{figma.com}), or even screenshots.
Generating dynamic UI involves a synthesis of state management logic, which requires an understanding of the render semantics of React~\citep{LeeAhnYi25React,MadLhoTip20Semantics}.
In many cases, generated code may be buggy and require substantial refinement by developers~\citep{AhmImr25Role,WeiCouLamDraMaa25AIInspired,PatGagSheThaVarShi25StarryStudioAI}.


\paragraph{{\normalfont\itshape Sketch-based Synthesis}}
To our knowledge, sketch-based synthesis~\citep{SolTanBodSesSar06Combinatorial} has not been applied to reactive UI synthesis.
For a static graphical output, \textsc{Sketch-n-Sketch}~\citep{ChuHemSprAlb16Programmatic,HemLubChu19SketchnSketch} has been developed to synthesize and manipulate programs by allowing users to directly interact with the output, utilizing sketch-based synthesis.

\paragraph{Reactive Synthesis from Examples}
Synthesizing reactive programs from examples is a relatively underexplored domain compared to reactive synthesis from temporal logic specifications.
\textsc{AutumnSynth}~\citep{DasTenSolTav23Combining} addresses this gap by proposing a two-layer approach that combines functional program synthesis and automata synthesis.
Unlike the grid-based setup using a simple DSL \textsc{Autumn} in \textsc{AutumnSynth}, \Rdemon{} generates stateful UIs written in JavaScript with React.

\textsc{Pasket}~\citep{JeoQiuFetFosSol16Synthesizing} synthesizes models of reactive frameworks like Swing and Android by taking logs of runtime events and API calls.
Notably, \textsc{Pasket} also utilizes a sketch-based approach during synthesis.

\section{Future Work}\label{sec:future}
\paragraph{{\normalfont\itshape Frontend Interface}}
\Rdemon{} currently works with mockups that may have been exported from UI prototyping tools like Figma.
Our \Sketch{} pane could be enhanced with direct manipulation~\citep{HemLubChu19SketchnSketch,ChuHemSprAlb16Programmatic} capabilities so that mockups can be built within \Rdemon{}.

For enhanced user experience, we plan to automatically record interactions without requiring manual submission of interactions.

\paragraph{Synthesis Backend}
Demo timelines could be used to automatically generate UI testing code, enabling verification that LLM-generated components adhere to the user specifications.

While the current backend simply falls back to using LLMs when the enumerative method fails, we are developing an approach to use the two methods in tandem, as explored in recent works~\citep{DinQiu24Enhanced,LiMagBraOBPol25Guided,PatRahMisBisDil24ProgrammingbyDemonstration}.

\bibliographystyle{ACM-Reference-Format}
\bibliography{references}
\end{document}